\documentclass[letterpaper,10pt]{article} 

\usepackage{opticameet3} 

\newcommand\authormark[1]{\textsuperscript{#1}}
\usepackage{subcaption,graphicx}
\usepackage{amsmath,amssymb}
\usepackage[colorlinks=true,bookmarks=false,citecolor=blue,urlcolor=blue]{hyperref} 

\begin{document}

\title{Temporal Multiplexing of Heralded Photons Based on Thin Film Lithium Niobate Photonics}


\author{Cagin Ekici,\authormark{1} Yonghe Yu,\authormark{1} Jeremy C. Adcock,\authormark{1} Alif Laila Muthali,\authormark{1} Mujtaba Zahidy,\authormark{1} Heyun Tan,\authormark{2} Zhongjin Lin,\authormark{3} Hao Li,\authormark{2}  Leif K. Oxenløwe,\authormark{1} Xinlun Cai,\authormark{2} and Yunhong Ding\authormark{1,*}}

\address{\authormark{1}Center for Silicon Photonics for Optical Communication (SPOC), Department of Electrical and Photonics Engineering, Technical University of Denmark, Lyngby, Denmark\\
\authormark{2}State Key Laboratory of Optoelectronic Materials and Technologies, School of Electronics and Information Technology, Sun Yat-sen University,
Guangzhou 510275, China\\
\authormark{3}Department of Electrical and Computer Engineering, The University of British Columbia, Vancouver, BC V6T 1Z4, Canada }

\email{\authormark{*}yudin@dtu.dk} 

\begin{abstract}
Heralded photons from a silicon source are temporally multiplexed utilizing thin film lithium niobate photonics. The time-multiplexed source, operating at a rate of R = 62.2 MHz, enhances single photon probability by 3.25 ± 0.05.
\end{abstract}

\section{Introduction}
Quantum photonics is emerging as a prominent candidate for the development and implementation of quantum technologies, manifesting its capabilities in demonstrating a quantum computational advantage \cite{madsen2022}, and ensuring error protection \cite{vigliar2021}. However, in order to fully harness the potential of quantum technologies beyond proof-of-principle experiments, periodic and near-deterministic single-photon sources are required.

Heralded photon sources, crucial in many quantum photonics investigations, generate a pair of single photons (signal and idler) through the virtual absorption of two pump photons, utilizing nonlinear optical processes, such as spontaneous four-wave mixing (SFWM). However, their performance is constrained by the limited probability of generating a photon pair, $p$, to avoid unwanted $k$-photon-pair generation, $\sim p^k$. This limitation impedes the scalability, making them unsuitable for more complex architectures. To address this challenge, a technique called multiplexing of probabilistic sources was proposed \cite{pittman2002,adcock2022qu}. The basic idea is to use a number of photon pair sources and exploit heralding (detection of idler photon) for routing the heralded photon (signal) to a single mode output, as depicted in Fig. \ref{multip} inset (a). Multiplexing can utilize various orthogonal degrees of freedom, including temporal, spatial or their combinations. Temporal multiplexing schemes, which require only a single photon detector, an actively controllable storage medium, and a photon pair source, offer a significant advantage over spatial multiplexing schemes \cite{collins2013} in terms of physical resources. Previous demonstrations of temporal multiplexing in fiber-based systems have been reported with enhancements of the single photon detection probability, E, up to 3 with R = 10 MHz \cite{xiong2016}  and E $= 4.5$ with R = 0.5 MHz \cite{adcock2022}, while free space implementation has achieved E $= 28$ with R = 0.5 MHz \cite{kaneda2019}. In this work, we present a temporal multiplexing scheme based on high-speed and low-loss thin film lithium niobate (TFLN) photonics operating at R = 62.2 MHz, with E $= 3.25 \pm 0.05$.

\section{Experimental Setup and Results}
\begin{figure}[!htbp]
\centering
\includegraphics[scale=0.92]{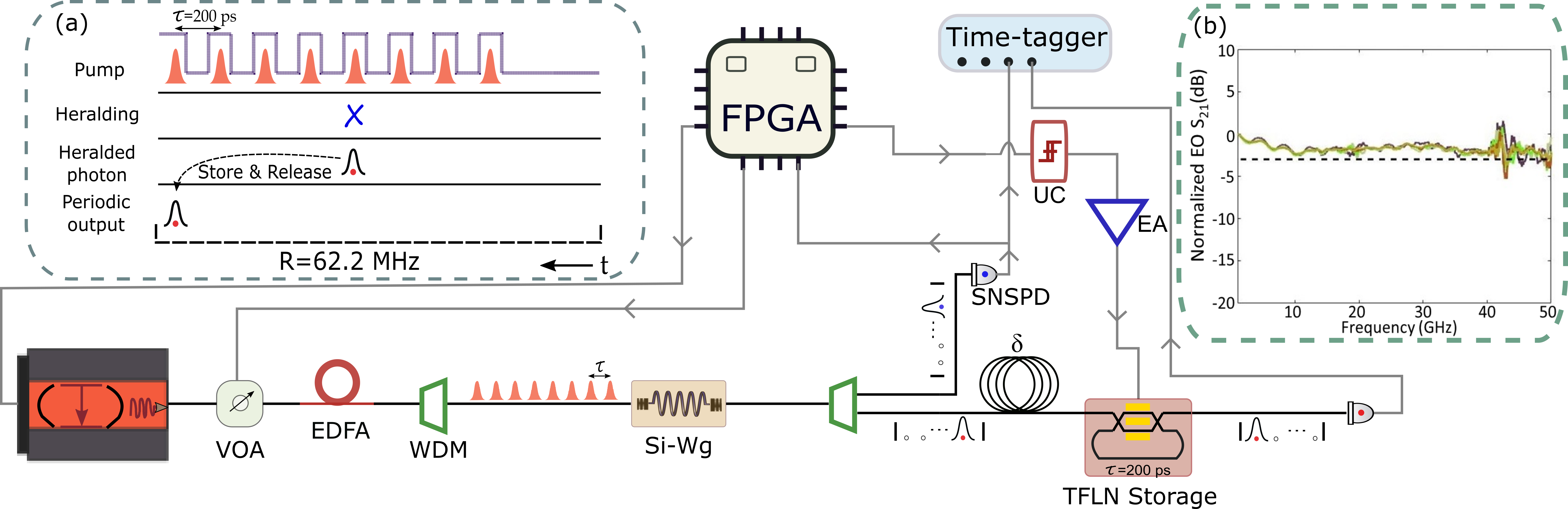}\vskip-0.9em
\caption{\label{multip} Schematic of the experimental setup with insets illustrating  (a) the basic idea of temporal multiplexing, including the values used in this experiment, and (b) electro-optic bandwidth (S$_{21}$) measurement results.}
\end{figure}

We multiplex 12 temporal modes within a single clock cycle using heralded photons generated in 1.2 cm long silicon waveguide source through SFWM. To trigger SFWM processes, 1545 nm bright pump pulses are subjected to a carving via a variable optical attenuator (VOA) to create temporally distinct photon pair sources and are amplified via an erbium-doped fiber amplifier (EDFA) before illuminating the waveguide. The temporal modes, with a spacing of 200 ps, are generated from a laser operating at 10 GHz repetition rate. This occurs every $\sim$ 16.07 ns, defining an output clock rate of 62.2 MHz (see Fig. \ref{multip}). The generated single photons are separated using off-chip WDM filters at energy-conserving wavelengths of 1540.56 nm and 1550.12 nm, defining herald (idler) and heralded (signal) photon channels, respectively. The herald photons are detected by a superconducting nanowire single photon detector (SNSPD) and recorded by a time-tagger to produce timing data of the detection events. The heralded photons are coupled to a high-resolution single photon buffer based on TFLN for multiplexing. The device, capable of resolving 200 ps time-steps, includes a 2.4 cm-long loop and a voltage-controlled interferometric switch, exhibiting a 3-dB electro-optic bandwidth exceeding 40 GHz (see Fig. \ref{multip} inset (b)) \cite{ekici2023}. The measured fiber-to-fiber insertion loss, including coupling, is below $-$6.2 dB.

\vskip-0.6em

Upon receiving a signal heralding the generation of a photon pair, the FPGA outputs a clocked pulse to control the TFLN switch. This pulse is then directed into an ultrafast comparator (UC) to achieve a fast fall-rise time, and is amplified via an electronic amplifier (EA) to match $V_\pi\: (\approx 6 \: \text{V})$ of the TFLN switch, before being applied to the device. This acts to store the partner photon until the predetermined output time bin, i.e., the final time bin defined by the pump laser pulse train. Synchronisation is achieved with a fiber delay of $\delta$ ns.

\begin{figure}[!htbp]
	\centering
	\begin{subfigure}{.48\linewidth}
		\centering
		\includegraphics[scale=0.156]{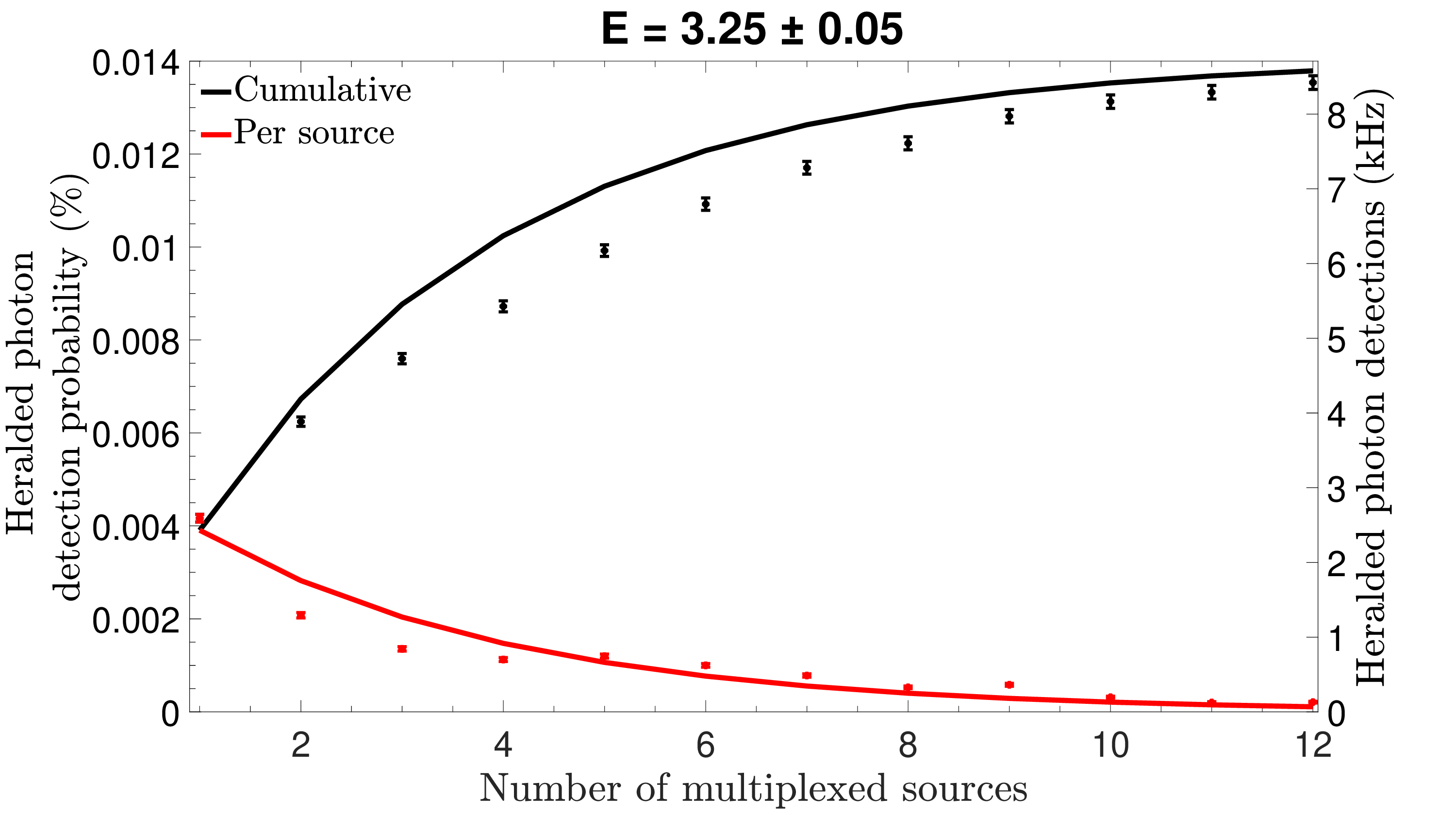}
		\caption{}
	\end{subfigure}%
\hskip0.105em
	\begin{subfigure}{.48\linewidth}
		\centering
	\includegraphics[scale=0.156]{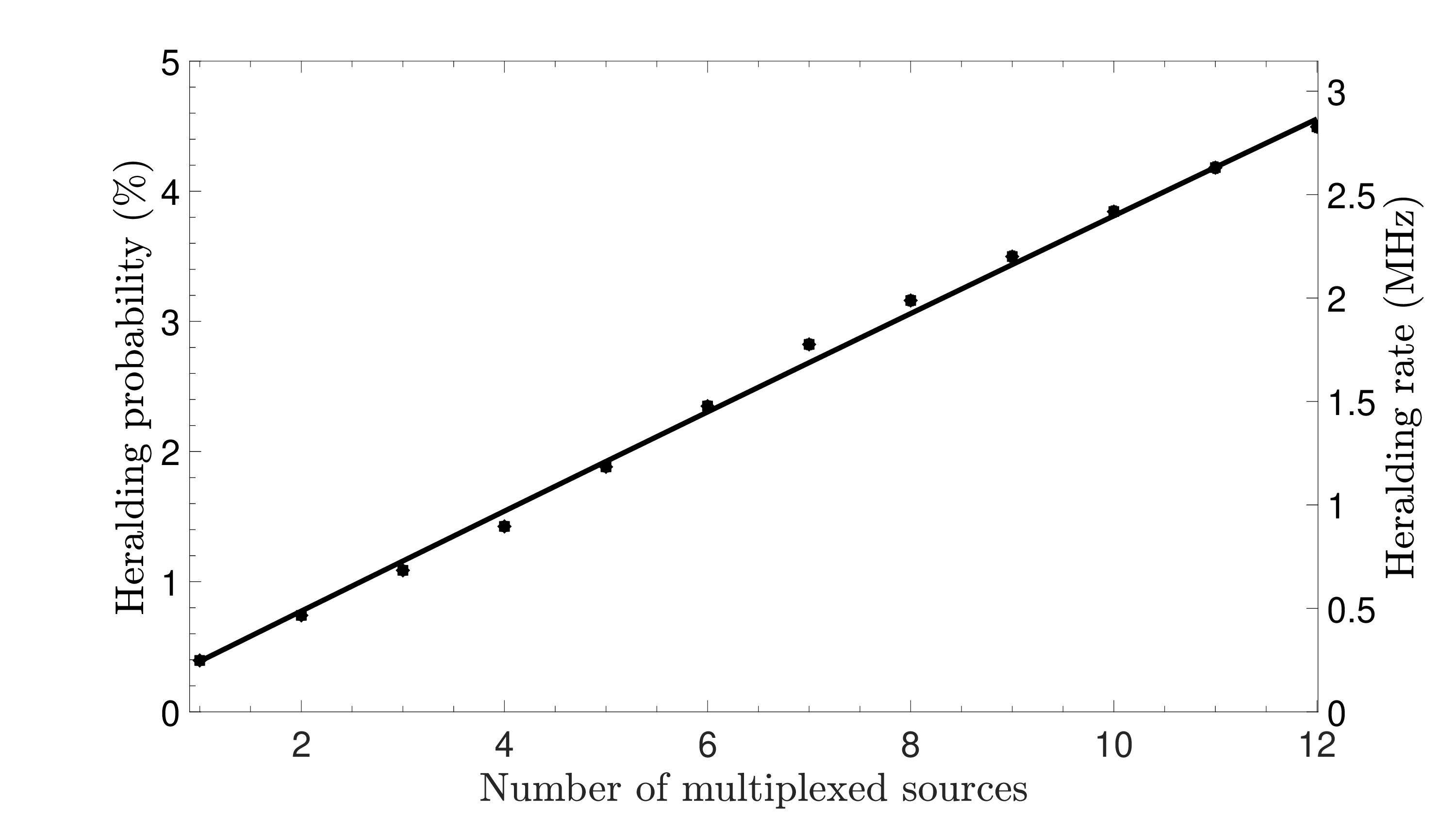}
		\caption{}
	\end{subfigure}%
\vskip-1.3em
	\caption{Temporal multiplexing experimental results including a numerical model (straight lines) \cite{kaneda2019}: (a) Heralded single photon rate and probability (b) Heralding detection rate and probability}
	 \label{enhan}%
\end{figure}

Fig. \ref{enhan} illustrates the measured probability versus the number of multiplexed sources (N). At a low pump power ($p\approx0.046$), the heralded photon detection probability is $p\eta_i\eta_s = 0.00004$ for the non-multiplexed source (N $= 1$) and $p_m = 0.00013$ for the multiplexed source (N $= 12$), resulting in E $= 3.25$, see Fig. \ref{enhan} (a). These values are observed for the transmission efficiencies of idler and signal photons, $\eta_i = 0.079$ and $\eta_s = 0.011$. The corresponding heralding rate and probability, depicted in Fig. \ref{enhan} (b), increase with the number of multiplexed sources.

\end{document}